\documentclass[aps,pra,twocolumn,showpacs,superscriptaddress,a4paper,groupedaddress]{revtex4-1}
\usepackage{graphicx}    
\usepackage{dcolumn}    
\usepackage{bm}             
\usepackage{amssymb}   
\usepackage{amsmath}    
\usepackage{subfigure}    
\usepackage{braket}
\usepackage{mathrsfs}
\usepackage[export]{adjustbox}
\usepackage[mathscr]{euscript}
\usepackage[toc,page]{appendix}
\begin{document}

\title{Degenerate Zeeman Ground States in the Single-Excitation Regime}
\author{R.T. Sutherland$^{1}$}
\email{rsutherl@purdue.edu}

\author{F. Robicheaux $^{1,2}$}
\email{robichf@purdue.edu}
\affiliation{$^{1}$Department of Physics and Astronomy, Purdue University, West Lafayette IN, 47907 USA}
\affiliation{$^{2}$Purdue Quantum Center, Purdue University, West Lafayette,
Indiana 47907, USA}

\date{\today}

\begin{abstract}
In this work, we demonstrate the importance of considering correlations between degenerate Zeeman sublevels that develop in dense atomic ensembles. In order to do this, we develop a set of equations capable of simulating large numbers of atoms while still incorporating correlations between degenerate Zeeman sublevels. This set of equations is exact in the single-photon limit, and may be interpreted as a generalization of the frequently used coupled harmonic oscillator equations. Using these equations, we demonstrate that in sufficiently dense systems, correlations between Zeeman sublevels can cause non-trivial differences in the photon scattering lineshape in arrays and clouds of atoms.
\end{abstract}
\pacs{42.50.Nn, 42.50.Ct, 32.70.Jz, 37.10.Jk}
\maketitle

\section{Introduction}\label{intro}
Since its inception \cite{dicke1954}, the collective optics of both ordered and disordered ensembles has been a vibrant field of study \cite{rouabah2014, bienaime2012, kaiser2015, bromley2016, pellegrino2014, javanainen2014, ruoskekoski2016, jennewein2016, browaeys2016_2, lee2016, svidzinsky2010, harde1991, gross1982, gross1976, gross1978, rehler1971, sutherland2016, sutherland2016_2, sutherland2017, bettles2015, bettles2016, scully2006, scully2007, scully2009, scully2009, scully2015, bender2010, courteille2010, kaiser2016_2, jen2017, jen2017_phase, ido2005, ruostekoski1997, hernandez2005, kuraptsev2014, monshouwer1997, palacios2002, zhu2015,zhu2016}. The low-excitation (single-photon) regime is of particular interest since, in this limit, exact computations of ensembles of thousands of $J_{g} = 0 \rightarrow J_{e}=1$ atoms are possible  \cite{rouabah2014, bienaime2012, kaiser2015, bromley2016, pellegrino2014, javanainen2014, ruoskekoski2016, jennewein2016, browaeys2016_2, lee2016, svidzinsky2010, sutherland2016, sutherland2016_2, bettles2015, bettles2016, bettles2017}. This allows theorists to accurately simulate large clouds and arrays of atoms \cite{sutherland2016, sutherland2016_2, jennewein2016, pellegrino2014}. Unfortunately, this technique only approaches exactness when the transitions probed have this angular momentum structure \cite{lee2016}. References \cite{bromley2016} and \cite{ido2005} are an example of such a system, where the absence of fine structure in $^{88}$Sr leads to a singlet ground state. However, most experimental setups do not contain this type of atom. For example, in many well-known experiments, $^{87}$Rb, which contains degenerate ground states because of its hyperfine structure, is the atom of choice \cite{roof2016, kaiser2016_2, jennewein2016, pellegrino2014}. This makes the development of potentially important quantum correlations possible. 

The difference between an ensemble of $J_{g}=0$ and $J_{g}\neq 0$ atoms could reasonably be mistaken as trivial, but it fundamentally changes the nature of the system. For atoms with $J_{g}=0$ and $J_{e} = 1$, no matter how large the cloud of atoms is, there is a single ground state: $\ket{ggg...g}$. Further, in the low-excitation regime, a calculation must contain $3N$ singly excited states, or, when there is significant magnetic splitting, $N$ singly excited states. This is not the case when $J_{g}\neq 0$, since the Hilbert space describing the set of ground states contains $(2J_{g} + 1)^{N}$ elements, and the set of singly excited states contains $N(2J_{e} + 1)(2J_{g}+1)^{N-1}$ elements. This makes the exact simulation of large clouds unfeasible, even in the single-photon limit. As a result, one approach that has been used is to stochastically distribute a Zeeman ground state, $M_{g}$, to each atom and set the correlation operators equal to zero \cite{pellegrino2014, jennewein2016, lee2016}, see Eq.~(\ref{eq:ruost_post}) below. As has been noted in Refs.~\cite{pellegrino2014,jennewein2016}, these calculations produce lineshapes that are qualitatively different than those observed experimentally.

It is well understood that quantum correlations can play an important role in sufficiently dense many-body systems. In various regimes, this is often dealt with by solving for a set of physically relevant operators. Simulations containing many degrees of freedom are then made possible by truncating the resulting hierarchy of equations, either by factorizing or dropping higher order correlation operators, via a cummulant expansion or some, physically meaningful, alternative argument \cite{kubo1962, zhu2015, holland2009, sutherland2016_2, lee2016}. The presence of degenerate Zeeman ground states introduces the potential for quantum correlations between different atoms and Zeeman states that are absent in the $J_{g}=0$ case.  When $J_{g}\neq 0$, i.e. there are multiply degenerate Zeeman ground states, the system behaves in a fundamentally quantum manner with correlations which, as will be shown below, change the optical properties of the system. It should be noted that this issue was originally addressed in Ref.~\cite{lee2016} through the field theoretical derivation of a hierarchy of operators capable of addressing these correlations to all orders. This present work derives a set of physically equivalent, yet perhaps more intuitive, equations by examining and exploiting the properties of the master equation (Eq.~(\ref{eq:full_master_2})) in the low excitation limit. Further, we illustrate the fundamental importance of addressing these correlations, by showing how the qualitative photon scattering lineshape in two experimentally relevant systems, strongly depends on the presence of quantum correlations.

Specifically, we will derive both the lower-order oscillator equations used in \cite{pellegrino2014, jennewein2016, lee2016}, followed by a generalized version of these equations. The latter of these sets of equations includes the correlations that occur when $J_{g} \neq 0$. Further, we will introduce an approximation to this generalized technique, capable of simulating systems with large values of $N$. We will then show that the two techniques give different photon scattering lineshapes in both atomic arrays and clouds. This indicates that in order to properly simulate clouds of $J_{g} \neq 0$ atoms, one must consider higher order quantum correlations. Lastly, we compare our results to the experimental observations of Ref.~\cite{jennewein2016}.

\section{Theory}
\subsection{Master Equation}

This work considers an ensemble of atoms interacting, via the dipole moments of an arbitrary $J_{g}\rightarrow J_{e}$ transition, with both a laser and the quantized electromagnetic field. Assuming the system is Markovian, we use the Lindblad form of the master equation for the reduced density matrix, $\hat{\rho}(t)$ \cite{james1993, agarwal2001}:
\begin{eqnarray}\label{eq:full_master_2}
\frac{d\hat{\rho}(t)}{dt} &=& -\frac{i}{\hbar}\Big[H,\hat{\rho}(t) \Big] \nonumber \\ &+& \Gamma\sum_{j}\sum_{M_{g}M_{g}^{\prime}}\sum_{q}C^{q}_{M_{g}}C^{q}_{M_{g}^{\prime}}\Big\{ \sigma^{j-}_{M_{g}q}\hat{\rho}(t)\sigma^{j+}_{M_{g}^{\prime}q}  \nonumber \\ &-& \frac{1}{2}\delta_{M_{g}M_{g}^{\prime}}\Big( \sigma^{j+}_{M_{g}q}\sigma^{j-}_{M_{g}^{\prime}q}\hat{\rho}(t) + \hat{\rho}(t)\sigma^{j+}_{M_{g}q}\sigma^{j-}_{M_{g}^{\prime}q}\Big)\Big\} \nonumber \\
&+&  \sum_{j\neq j^{\prime}}\sum_{M_{g}M_{g}^{\prime}}\sum_{qq^{\prime}}  C^{q}_{M_{g}}C^{q^{\prime}}_{M_{g}^{\prime}} \Big( \Big\{ g^{qq^{\prime}}_{jj^{\prime}}(\alpha_{jj^{\prime}}) \nonumber \\ &+& g^{qq^{\prime}}_{jj^{\prime}}(-\alpha_{jj^{\prime}})\Big\}\sigma^{j^{\prime} -}_{M_{g}^{\prime}q^{\prime}}\hat{\rho}(t)\sigma^{j^{+}}_{M_{g}q} \nonumber \\ &-& g^{qq^{\prime}}_{jj^{\prime}}(\alpha_{jj^{\prime}})\sigma^{j +}_{M_{g}q}\sigma^{j^{\prime} -}_{M_{g}^{\prime}q^{\prime}}\hat{\rho}(t)  \nonumber \\ &-&  g^{qq^{\prime}}_{jj^{\prime}}(-\alpha_{jj^{\prime}})\hat{\rho}(t)\sigma^{j +}_{M_{g}q}\sigma^{j^{\prime -}}_{M_{g}^{\prime}q^{\prime}} \Big).
\end{eqnarray}
Here $C^{q}_{M_{g}}\equiv \langle{J_{e} (M_{g}+q)}|1q;J_{g}M_{g}\rangle$ is a Clebsch-Gordon coefficient, $\sigma^{j +}_{M_{g}q}\equiv \ket{e J_{e}(M_{g} + q)}_{j}\bra{g J_{g} M_{g}} = (\sigma^{j -}_{M_{g}q})^{\dagger}$, $\Gamma$ is the single atom decay rate, $\alpha_{jj^{\prime}} \equiv k|\boldsymbol{r}_{j} - \boldsymbol{r}_{j^{\prime}}|$, and $g^{qq^{\prime}}_{jj^{\prime}}(\alpha_{jj^{\prime}})$ is the inelastic and elastic dipole-dipole Green's function:
\begin{eqnarray}
g^{qq^{\prime}}_{jj^{\prime}}(\alpha_{jj^{\prime}}) &\equiv & \nonumber \frac{3\Gamma e^{i\alpha_{jj^{\prime}}}}{4\alpha_{jj^{\prime}}}\Bigg[ i\Big\{ (\hat{\boldsymbol{e}}^{*}_{q}\cdot \hat{\boldsymbol{n}}_{jj^{\prime}})(\hat{\boldsymbol{e}}_{q^{\prime}}\cdot \hat{\boldsymbol{n}}_{jj^{\prime}}) - \delta_{qq^{\prime}}\Big\} \\ + \Big(\frac{1}{\alpha_{jj^{\prime}}} &+& \frac{i}{\alpha_{jj^{\prime}}^{2}} \Big) \Big\{\delta_{qq^{\prime}} - 3(\hat{\boldsymbol{e}}^{*}_{q}\cdot \hat{\boldsymbol{n}}_{jj^{\prime}})(\hat{\boldsymbol{e}}_{q^{\prime}}\cdot \hat{\boldsymbol{n}}_{jj^{\prime}}) \Big\}\Bigg], \nonumber \\ 
\end{eqnarray}
where $\hat{\boldsymbol{n}}_{jj^{\prime}} \equiv (\boldsymbol{r}_{j}-\boldsymbol{r}_{j^{\prime}})/|\boldsymbol{r}_{j} - \boldsymbol{r}_{j^{\prime}}|$, and $\hat{\boldsymbol{e}}_{q}$ represents the $q^{th}$ unit polarization vector. Using the rotating-wave approximation, 
\begin{eqnarray}
H &=& -\hbar\Delta\sum_{j}\sum_{M_{e}} \ket{eJ_{e}M_{e}}_{j}\bra{eJ_{e}M_{e}}\nonumber \\ &+& \frac{\mathcal{D}}{2}\sum_{j}\sum_{M_{g}}\sum_{q}C^{q}_{M_{g}} \Big\{ (\hat{\boldsymbol{e}}^{*}_{q}\cdot \boldsymbol{E}_{j})\sigma^{j +}_{M_{g}q} +  c.c.\Big\},
\end{eqnarray}
where $\Delta$ is the detuning of the laser, $\mathcal{D}$ is the reduced dipole matrix element \cite{shankar2012}, and $\boldsymbol{E}_{j}$ is the laser field for atom $j$.

\subsection{Hierarchy of Correlation Functions}

In order to simulate such highly correlated systems, in this section, we will derive a hierarchy of operator expectation values using Eq.~(\ref{eq:full_master_2}). Note that a physically equivalent, field theoretical, set of equations is presented in Ref.~\cite{lee2016}. Here, we calculate the time dependence of the expectation value of the $\sigma^{j -}_{M_{g}q}$ operators using the equation:
\begin{equation}
\frac{d\langle \sigma^{j -}_{M_{g}q} \rangle}{dt} = Tr\Big\{\sigma^{j -}_{M_{g}q} \frac{d\hat{\rho}(t)}{dt} \Big\}.
\end{equation}
This gives:
\begin{eqnarray}\label{eq:full_sigma}
\frac{d\langle \sigma^{j -}_{M_{g}q}\rangle}{dt} &=& (i\Delta -\Gamma/2)\langle \sigma^{j -}_{M_{g}q} \rangle \nonumber  \\ 
&-& \nonumber \frac{i\mathcal{D}}{2\hbar} \sum_{M_{g}^{\prime}}\sum_{q^{\prime}} (\hat{\boldsymbol{e}}^{*}_{q^{\prime}}\cdot \boldsymbol{E}_{j})C^{q^{\prime}}_{M_{g}^{\prime}}\Big\{ \\ 
 && \langle \sigma^{j -}_{M_{g}q}\sigma^{j +}_{M_{g}^{\prime}q^{\prime}} \rangle\delta_{_{(M_{g} +q)(M_{g}^{\prime}+q^{\prime})}}  \nonumber
- \langle \sigma^{j +}_{M_{g}^{\prime}q^{\prime}}\sigma^{j -}_{M_{g}q} \rangle \delta_{_{M_{g}M_{g}^{\prime}}}  \Big\} \\
 &-& \nonumber \sum_{j^{\prime\prime}\neq j}\sum_{M_{g}^{\prime}M_{g}^{\prime\prime}}\sum_{q^{\prime}q^{\prime\prime}}C^{q^{\prime}}_{M_{g}^{\prime}}C^{q^{\prime\prime}}_{M_{g}^{\prime\prime}} g^{q^{\prime}q^{\prime\prime}}_{jj^{\prime\prime}}(\alpha_{jj^{\prime\prime}})\Big\{  \\ &&\langle \sigma^{j^{\prime\prime} -}_{M_{g}^{\prime\prime}q^{\prime\prime}}\sigma^{j -}_{M_{g}q}\sigma^{j +}_{M_{g}^{\prime}q^{\prime}} \rangle \delta_{_{(M_{g}+q)(M_{g}^{\prime} + q^{\prime})}} \nonumber \\
 &-& \langle \sigma^{j^{\prime\prime} -}_{M_{g}^{\prime\prime}q^{\prime\prime}}\sigma^{j +}_{M_{g}^{\prime}q^{\prime}}\sigma^{j -}_{M_{g}q} \rangle \delta_{_{M_{g}M_{g}^{\prime}}}  \Big\}.
\end{eqnarray}
For this system, it is useful to probe the low-intensity ($|E||\mathcal{D}| \ll \Gamma\hbar$) limit. This allows one to set terms $\propto \sigma^{j +}_{M_{g}^{\prime}q^{\prime}}\sigma^{j -}_{M_{g}q}$ equal to zero, since these terms only project onto states with excited atoms. Further, we consider only the case where the system is \textit{initially} in an uncorrelated mixed state. This may be simulated by stochastically assigning each atom a Zeeman ground state, $M_{g}^{j}$, and then averaging over many runs \cite{pellegrino2014, lee2016}. This allows us to make the approximation $\langle \sigma^{j -}_{M_{g}^{\prime}q^{\prime}}\sigma^{j +}_{M_{g}q}\rangle \rightarrow \delta_{M_{g}^{\prime}M_{g}}\delta_{qq^{\prime}}\delta_{M_{g}M^{j}_{g}}$, since the developement of correlations between different $M_{g}$ values, for states with only ground state atoms, is increasingly slow in the low-intensity limit. Considering these approximations gives:
\begin{eqnarray}\label{eq:ruost_pre}
\frac{d\langle \sigma^{j -}_{M_{g}q}\rangle}{dt} &\simeq & (i\Delta -\Gamma/2)\langle \sigma^{j -}_{M_{g}q} \rangle - \frac{i\mathcal{D}}{2\hbar}(\hat{\boldsymbol{e}}^{*}_{q} \cdot \boldsymbol{E}_{j})C^{q}_{M_{g}}\delta_{_{M_{g}M_{g}^{j}}} \nonumber \\
 &-& \nonumber  \sum_{j^{\prime\prime}\neq j}\sum_{M_{g}^{\prime}M_{g}^{\prime\prime}}\sum_{q^{\prime}q^{\prime\prime}}C^{q^{\prime}}_{M_{g}^{\prime}}C^{q^{\prime\prime}}_{M_{g}^{\prime\prime}} g^{q^{\prime}q^{\prime\prime}}_{jj^{\prime\prime}}(\alpha_{jj^{\prime\prime}})  \nonumber \Big\{ \\
 && \langle \sigma^{j^{\prime\prime} -}_{M_{g}^{\prime\prime}q^{\prime\prime}}\sigma^{j -}_{M_{g}q}\sigma^{j +}_{M_{g}^{\prime}q^{\prime}} \rangle\delta_{_{(M_{g}+q)(M_{g}^{\prime}+q^{\prime})}} \Big\}.
\end{eqnarray}
In order to solve for the one atom correlation operators, one must solve for the two atom correlation operators, then the three atom correlation operators, etc... This produces a hierarchy of equations that grows exponentially with $N$. Because of this, the expectation values of the correlation functions are often factorized in the following way:
\begin{eqnarray}\label{eq:fact}
\langle \sigma^{j^{\prime\prime} -}_{M_{g}^{\prime\prime}q^{\prime\prime}}\sigma^{j -}_{M_{g}q}\sigma^{j +}_{M_{g}^{\prime}q^{\prime}} \rangle &\simeq & \langle \sigma^{j^{\prime\prime} -}_{M_{g}^{\prime\prime}q^{\prime\prime}}\rangle \langle \sigma^{j -}_{M_{g}q}\sigma^{j +}_{M_{g}^{\prime}q^{\prime}} \rangle \nonumber \\
&& \simeq  \langle \sigma^{j^{\prime\prime} -}_{M_{g}^{\prime\prime}q^{\prime\prime}}\rangle \delta_{_{M_{g}M_{g}^{j}}}\delta_{_{M_{g}M_{g}^{\prime}}}\delta_{_{qq^{\prime}}},
\end{eqnarray}
keeping in mind that we have stochastically assigned a value of $M_{g}^{j}$ to each atom. This gives:
\begin{eqnarray}\label{eq:ruost_post}
\frac{d\langle \sigma^{j -}_{M_{g}q}\rangle}{dt} &\simeq & (i\Delta -\Gamma/2)\langle \sigma^{j -}_{M_{g}q} \rangle - \frac{i\mathcal{D}}{2\hbar}(\hat{\boldsymbol{e}}^{*}_{q} \cdot \boldsymbol{E}_{j})C^{q}_{M_{g}}\delta_{_{M_{g}M_{g}^{j}}} \nonumber \\
 && - \nonumber  \sum_{j^{\prime\prime}\neq j}\sum_{M_{g}^{\prime}}\sum_{q^{\prime}}C^{q}_{M_{g}}C^{q^{\prime}}_{M_{g}^{\prime}} g^{qq^{\prime}}_{jj^{\prime\prime}}(\alpha_{jj^{\prime\prime}}) \langle \sigma^{j^{\prime\prime} -}_{M_{g}^{\prime}q^{\prime}}\rangle \delta_{M_{g}M_{g}^{j}}. \nonumber \\
\end{eqnarray}
The approximate equations above are the ones typically used \cite{jennewein2016,sutherland2016,sutherland2016_2,pellegrino2014,bromley2016}, and are exact when $J_{g} = 0$ and $J_{e} = 1$ \cite{jennewein2016}. The factorized terms in Eq.~(\ref{eq:fact}) are often \textit{highly} correlated for dense systems, making this approximation inadequate. In fact, as will be demonstrated in Sec.~\ref{sec:results}, their presence changes the system's photon scattering in non-trivial ways. A technique capable of addressing these correlations, while remaining computationally pragmatic, is required.

\subsection{Reduced Master Equation}

In this section, we develop a system of equations capable of addressing the correlations that emerge in Eq.~(\ref{eq:full_master_2}), in the low-intensity limit. An alternate, yet physically equivalent, set of equations is presented in Ref.~\cite{lee2016}. First, we write the density matrix in the following way:
\begin{equation}
\hat{\rho}(t) = \sum_{e,e^{\prime}}\sum_{m,n} c_{e,m; e^{\prime},n}(t)\ket{e,m}\bra{e^{\prime},n}.
\end{equation}
Here, $e$ represents the number of excitations in a given state, and $m$ is an index that runs over all the states within that subspace. We assume that the system is initialized in the pure state $\ket{0,g}$ and subsequently driven by a low-intensity laser. In the low-intensity limit:
\begin{eqnarray}
|c_{0,g;0,g}| \simeq 1 \gg |c_{1,m;0,g}| \gg |c_{1,m^{\prime};1,n^{\prime}}|.
\end{eqnarray}
We also note that the rate of population transfer is $\propto c_{1,m;1,n}\Gamma$, while the system approaches a quasi-steady state at a rate $\propto\Gamma$. Therefore, we neglect the effects of population transfer, i.e. terms in Eq.~(\ref{eq:full_master_2}) $\propto \sigma^{j -}_{M_{g}q}\hat{\rho}(t)\sigma^{j^{\prime} +}_{M_{g}^{\prime}q^{\prime}}$. Considering this, Eq.~(\ref{eq:full_master_2}) shows that the $c_{0,g;0,g}$ density matrix element only couples to the $c_{1,m;0,g}$ elements, as well as their complex conjugates. Also, to first order, the $c_{1,m;0,g}$ density matrix elements are only coupled to the $c_{0,g;0,g}$, as well as the $c_{1,n;0,g}$ elements. The relevant space of density matrix elements only couples to states with $\bra{0,g}$ as the column state, therefore, the rows and the columns of the density matrix are completely uncorrelated, allowing us to rewrite $\hat{\rho}(t)$ as:
\begin{equation}
\hat{\rho}(t) \simeq \sum_{e,e^{\prime}}\sum_{m,n} c_{e,m}(t)c_{e^{\prime},n}^{*}(t)\ket{e,m}\bra{e^{\prime},n},
\end{equation}
dramatically reducing the number of independent complex numbers needed to describe the density matrix. Further, in the low-intensity limit, the following properties are true:
\begin{eqnarray}
|c_{0,g}(t)|^{2} &\simeq & 1 \nonumber \\ 
c_{1,m}(t)c^{*}_{0,g}(t) &\simeq & c_{1,m}(t) \nonumber \\ 
|c_{1,m}(t)c^{*}_{0,g}(t)|  &\gg & |c_{1,m}(t)c^{*}_{1,k}(t)| \nonumber \\ 
|c_{1,m}(t)c^{*}_{1,k}(t)| &\gg & \mathcal{O}\Big\{\Big(\frac{\mathcal{D}|E|}{2\hbar \Gamma}\Big)^{3}\Big\}.
\end{eqnarray}
Considering the above approximations:
\begin{equation}
\Big( c_{1,m}(t)c^{*}_{0,g}(t)\Big) \Big(c_{0,g}(t)c^{*}_{1,n}(t)\Big) \simeq c_{1,m}(t)c^{*}_{1,n}(t),
\end{equation}
which allows one to construct any physical quantities $\propto c_{1m}(t)c^{*}_{1,n}(t)$, such as the photon scattering rate, from the significantly smaller subspace of $c_{1,m}(t)c^{*}_{0,g}(t) \simeq c_{1,m}(t)$ density matrix elements. Note that the $c_{1,m}(t)c_{0,g}^{*}(t)$ terms will be labeled simply $c_{1,m}(t)$ from now on. Resultantly, we may solve for the system's dynamics from the $c_{1,m}(t)$ subspace, tracking only the operations on the left side of $\hat{\rho}(t)$. This gives:
\begin{eqnarray}\label{eq:main_eq_low}
\frac{dc_{1,m}(t)}{dt} &= \nonumber & (i\Delta - \Gamma/2)c_{1,m}(t) - \frac{i\mathcal{D}}{2\hbar}(\hat{\boldsymbol{e}}^{*}_{q} \cdot \boldsymbol{E}_{j})C^{q}_{M_{g}}\delta_{_{M_{g}M_{g}^{j}}} \\ 
&-&  \sum_{n\neq m}\sum_{jj^{\prime}}\sum_{M_{g}M_{g}^{\prime}}\sum_{qq^{\prime}}C^{q}_{M_{g}}C^{q^{\prime}}_{M_{g}^{\prime}}\Big\{ \nonumber \\ && g^{qq^{\prime}}_{jj^{\prime}}(\alpha_{jj^{\prime}})\bra{1,n}\sigma^{j +}_{M_{g}q}\sigma^{j^{\prime} -}_{M_{g}^{\prime}q^{\prime}}\ket{1,m}c_{1,n}(t) \Big\}. \nonumber \\
\end{eqnarray}
The amount of elements in this subspace is $N(2J_{e}+1)(2J_{g}+1)^{N-1}$, which is considerably less than the $4^{N}(J_{e} + J_{g} + 1)^{2N}$ elements in the full density matrix. This allows us to simulate clouds of significantly more atoms, while still including the relevant correlations present in systems with $J_{g} \neq 0$. An interesting feature of Eq.~(\ref{eq:main_eq_low}) is that it is equivalent to Eq.~(\ref{eq:ruost_post}) for ensembles of $J_{g} = 0 \rightarrow J_{e} = 1$ atoms. For this angular momentum structure, both approaches are exact in the low-intensity regime, and are equivalent to the coupled harmonic oscillator approach typically used in the literature \cite{rouabah2014, bienaime2012, kaiser2015, bromley2016, pellegrino2014, javanainen2014, ruoskekoski2016, jennewein2016, browaeys2016_2, lee2016, svidzinsky2010, sutherland2016, sutherland2016_2, bettles2015, bettles2016, bettles2017}. This indicates that Eq.~(\ref{eq:main_eq_low}) may be viewed as a generalization to these equations.

\begin{figure}[!tp]
	\centering
	\includegraphics[width=0.4\textwidth,scale=0.1]{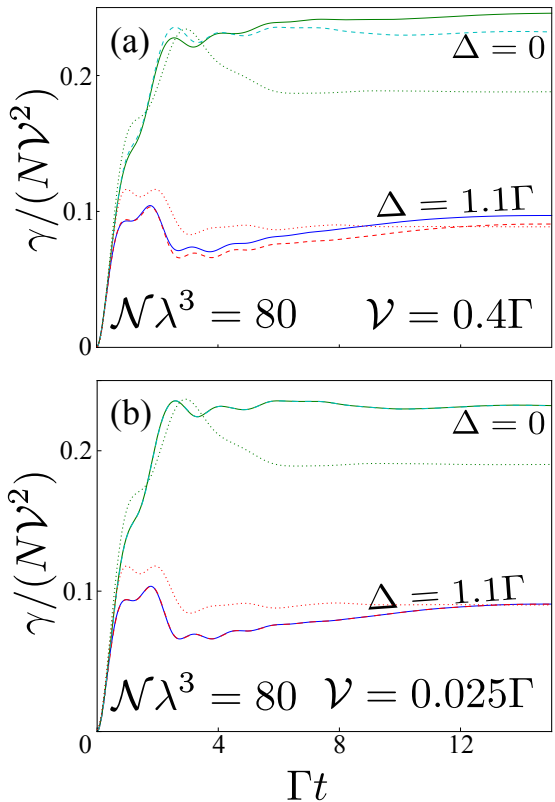}
    \caption{Comparison between the photon scattering rate, $\gamma$, calculated using Eq.~(\ref{eq:full_master_2}) (solid lines), Eq.~(\ref{eq:ruost_post}) (dotted lines), and the approximate Eq.~(\ref{eq:main_eq_low}) (dashed lines), for a single, dense, randomly distributed, Gaussian ensemble of 4 atoms. (a) Compares the photon scattering rate for systems being driven with a laser such that $\mathcal{V} = 0.4\Gamma$, where $\mathcal{V}\equiv |E_{z}||\mathcal{D}|/2\hbar$. This is shown for both on-resonance light, $\Delta = 0$, and blue-detuned light $\Delta = 1.1\Gamma$. (b) Shows the same comparison for $\mathcal{V} = 0.025\Gamma$. This demonstrates that the two calculations give the same answer as $\mathcal{V}/\Gamma\rightarrow 0$. Note that the discrepancies between calculations using Eq.~(\ref{eq:ruost_post}) and using Eq.~(\ref{eq:main_eq_low}) do not diminish with laser intensity.}
    \label{fig:photcomp}
\end{figure}

\begin{figure}[!tp]
	\centering
	\includegraphics[width=0.4\textwidth,scale=0.1]{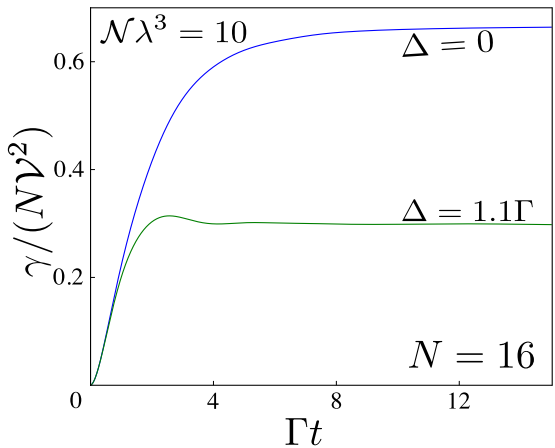}
    \caption{In this figure, we demonstrate the photon scattering, $\gamma$, for a single stochastically generated cloud of $16$ $J_{g} = 1/2 \rightarrow J_{e} = 1/2$ atoms versus time.}
    	\label{fig:show_off}
\end{figure}

Equation~(\ref{eq:main_eq_low}) approaches exactness in the limit where the system scatters a single real photon. This is shown in Fig.~\ref{fig:photcomp}. Here, we compare calculations of the photon scattering rate, $\gamma$, for individual configurations of dense clouds of $4$ $J_{g} = 1/2$, $J_{e} = 1/2$ atoms driven by a plane wave laser. In this figure, it is shown that for larger Rabi frequencies, the photon scattering rate obtained using Eq.~(\ref{eq:full_master_2}) yields different results than Eq.~(\ref{eq:main_eq_low}). However, for lower intensity lasers the two calculations are equivalent. Fig.~\ref{fig:photcomp} also demonstrates that simulations using Eq.~(\ref{eq:ruost_post}) and Eq.~(\ref{eq:main_eq_low}) produce different photon scattering rates, even at low laser intensities. Note that the results from Eq.~(\ref{eq:main_eq_low}) approach the exact result (i.e. Eq.~(\ref{eq:full_master_2})) while Eq.~(\ref{eq:ruost_post}) does not. Equation~(\ref{eq:main_eq_low}) has the capacity to simulate much larger ensembles than Eq.~(\ref{eq:full_master_2}). This is demonstrated in Fig.~\ref{fig:show_off}, where $\gamma$ for a cloud of $16$ atoms is calculated. Simulations with this many atoms are not possible when using Eq.~(\ref{eq:full_master_2}), since the number of elements in the density matrix is more than $10^{13}$ times larger than the number required to use Eq.~(\ref{eq:main_eq_low}).

\subsection{First-order Correlations}

While Eq.~(\ref{eq:main_eq_low}) allows one to simulate clouds containing significantly more atoms than Eq.~(\ref{eq:full_master_2}) does, its computational capacity falls short of the ability to simulate clouds containing hundreds of atoms. Since this regime is that of recent experiments \cite{pellegrino2014, jennewein2016}, an algorithm that can both simulate hundreds of atoms, and still keep the \textit{most relevant} correlations, is needed. In order to accomplish this, we truncate the set of terms that Eq.~(\ref{eq:main_eq_low}) spans.

First, the system is stochastically assigned the pure state with $0$ excitations, labeled $\ket{0,g}$. Here, and in the rest of this work, we assume a $\hat{\boldsymbol{z}}$ polarized laser. Upon a given initialization, the set of $c_{1,n}\ket{1,n}\bra{0,g}$ obtained through all permutations of the following operation is kept:
\begin{eqnarray}\label{eq:closed}
\sigma^{j^{\prime} +}_{M_{g}^{\prime\prime}q^{\prime\prime}}\sigma^{j -}_{M_{g}^{\prime}q^{\prime}}\sigma^{j +}_{M_{g}0}\ket{0,g},
\end{eqnarray}
while all other terms are set to $0$. In words, we keep the terms that are obtainable through an absorption of a single $\hat{\boldsymbol{z}}$ polarized photon from the laser, as well as the states obtainable through the exchange of one photon between atoms. This is useful because the size of the relevant space scales $\propto N^{2}$ as opposed to exponentially. Unlike Eq.~(\ref{eq:main_eq_low}), this algorithm's scaling with $N$ is independent of the level-structure of the system, allowing for the simulation of large clouds of atoms with complex angular momentum structures. The following sections demonstrate that these approximations yield accurate results in many relevant systems.

\section{Results}\label{sec:results}

\subsection{Photon Scattering in an Array of Atoms}\label{sec:line}

The collective nature of dipole-dipole interactions in a one dimensional array has been shown to produce non-trivial radiative properties \cite{sutherland2016_2,bettles2017,jennewein2016, jen2017_phase}. For this system, we show the importance of correlations between degenerate Zeeman ground states by demonstrating their effect on the photon scattering of an array containing $J_{g} = 1/2 \rightarrow J_{e} = 1/2$ atoms. Here the array is aligned along the $\hat{\boldsymbol{y}}$ direction, while the laser is a plane wave propagating along the $\hat{\boldsymbol{x}}$ direction and polarized in the $\hat{\boldsymbol{z}}$ direction. All of the results here are for an initially uncorrelated mixuture of Zeeman ground states, where every value of $M_{g}$ has an equal probability. For all the calculations shown, this is simulated by stochastically averaging over 320 runs where every atom is assigned a random value of $M_{g}$.

\begin{figure}[!tp]
	\centering
	\includegraphics[width=0.4\textwidth,scale=0.1]{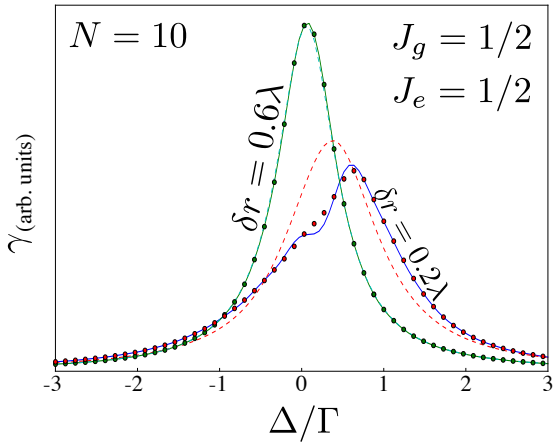}
    \caption{The photon scattering, $\gamma$, vesus $\Delta$ for 1D arrays of 10 $J_{g}=1/2 \rightarrow J_{e}=1/2$ atoms. The lineshape is calculated using Eq.~(\ref{eq:ruost_post}) (dotted lines), Eq.~(\ref{eq:closed}) (solid circles), and Eq.~(\ref{eq:main_eq_low}) (solid lines) for various array spacings, $\delta r$. Note that for larger values of $\delta r$, the two calculations agree well, indicating that correlations between degenerate Zeeman states are less important. However, for smaller array spacings the effects of correlations on the photon scattering are clearly seen.}
    	\label{fig:ruolinecompare}
\end{figure}

The results from Eq.~(\ref{eq:main_eq_low}) are compared with the results from both Eq.~(\ref{eq:ruost_post}) and Eq.~(\ref{eq:closed}). This is, to some extent, a good quantification of the importance of correlations in the photon scattering process. Equation~(\ref{eq:main_eq_low}) contains correlations between Zeeman states to all orders, Eq.~(\ref{eq:closed}) only includes the correlations resulting from a single virtual photon exchange between states, while Eq.~(\ref{eq:ruost_post}) contains none of these correlations.

Figure~\ref{fig:ruolinecompare} compares the steady-state photon scattering rate, $\gamma$, versus detuning $\Delta$ for an array of atoms using Eq.~(\ref{eq:ruost_post}), Eq.~(\ref{eq:closed}), and Eq.~(\ref{eq:main_eq_low}). In the figure, it is seen that for array spacings of $0.6\lambda$, the photon scattering lineshape is equivalent using all three methods. This indicates that in this regime, higher order correlations do not play a significant role in the system. Since the rate of virtual photon exchanges between states is $\propto 1/\delta r^{3}$, for smaller array spacings, correlations develop and the photon scattering lineshapes given by the three methods become qualitatively different. This is shown in Fig.~\ref{fig:ruolinecompare}, where for array spacings of $\delta r = 0.2\lambda$, the lineshape calculated using Eq.~(\ref{eq:ruost_post}) is still a Lorentzian structure, while the lineshapes calculated using both Eq.~(\ref{eq:main_eq_low}) and Eq.~(\ref{eq:closed}) have deviated from a Lorentzian shape, clearly forming double peaked structures. The figure demonstrates that for this value of $\delta r$, many of the differences between Eq.~(\ref{eq:ruost_post}) and Eq.~(\ref{eq:main_eq_low}) can be explained by the first-order correlations given by Eq.~(\ref{eq:closed}), since two the calculations produce similar results. For even smaller array spacings, however, the results between all three calculations qualitatively differ. This is indicative of the fact that, here, even higher order correlations need to be included. 

\subsection{Photon Scattering in a Cloud of Atoms}\label{sec:cloud}

In this section, we calculate the lineshape of the scattered radiation from a Gaussian cloud, with an average density $\mathcal{N} =\frac{N}{\sigma_{x}\sigma_{y}\sigma_{z}(4\pi)^{3/2}}$ and a distribution:
\begin{equation}
\rho(\boldsymbol{r}) = \frac{N}{\sigma_{x}\sigma_{y}\sigma_{z}(2\pi)^{3/2}}\exp\Big(-\Big\{\frac{x^{2}}{2\sigma_{x}^{2}} + \frac{y^{2}}{2\sigma_{y}^{2}} + \frac{z^{2}}{2\sigma_{z}^{2}}\Big\} \Big),
\end{equation}
driven by a plane wave laser polarized in the $\hat{\boldsymbol{z}}$ direction. This system is of particular interest due to recent experiments on cold clouds of $^{87}$Rb \cite{pellegrino2014,jennewein2016}. $^{87}$Rb has two hyperfine ground states: $(5S_{1/2} F=1)$ and $(5S_{1/2} F=2)$.  In these two experiments, the atoms are optically pumped into the $(5S_{1/2} F=2)$ ground state, and the closed transition to the $(5P_{3/2} F=3)$ excited state is probed. Presently, we demonstrate that such correlations make non-trivial differences in the optical properties of clouds. We should note that this section is incomplete with respect to the task of simulating the experimental results of Refs.~\cite{pellegrino2014,jennewein2016}, since it assumes, among other things, that the initial ground state is a completely uncorrelated mixture of Zeeman sublevels. However, it does demonstrate that these correlations play a vital role in understanding the lineshape of a cloud. All of the results here are for an initially uncorrelated mixuture of Zeeman ground states, where every value of $M_{g}$ has an equal probability of being populated. For all the calculations shown, this is simulated by stochastically averaging over $6000/N$ runs where every atom is assigned a random position and value of $M_{g}$.

\begin{figure}[!tp]
	\centering
	\includegraphics[width=0.4\textwidth,scale=0.1]{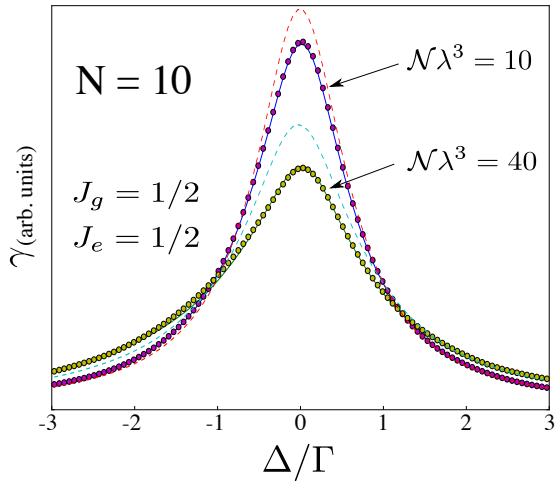}
    \caption{The photon scattering for spherically symmetric gaussian clouds of $10$, $J_{g} = 1/2 \rightarrow J_{e} = 1/2$ atoms versus detuning, $\Delta$, for various densities $\mathcal{N}$. The importance of correlations between degenerate Zeeman ground states is shown by comparing Eq.~(\ref{eq:main_eq_low}) (solid lines), Eq.~(\ref{eq:closed}) (solid circles), and Eq.~(\ref{eq:ruost_post}) (dotted lines). Note that the figure shows reasonable aggreement between the results of Eq.~(\ref{eq:main_eq_low}) and Eq.~(\ref{eq:closed}). This shows that for the values of $\mathcal{N}$ shown, most of the relevant physics is described by first order correlations.}
	\label{fig:gausscompareclosed}    
\end{figure}

To start, Fig.~\ref{fig:gausscompareclosed} shows the importance of correlations for spherically symmetric clouds of $10$ $J_{g} = 1/2 \rightarrow J_{e} = 1/2$ atoms. Here all three calculations have been compared for different values of $\mathcal{N}\lambda^{3}$. Again, one can see that Eq.~(\ref{eq:ruost_post}) is insufficient to describe the photon scattering of the system. While Eq.~(\ref{eq:main_eq_low}) allows for the simulation of clouds of atoms that were previously unreachable, experimental relevance still requires significantly more atoms. Fortunately, Fig.~\ref{fig:gausscompareclosed} shows that for values of $\mathcal{N}\lambda^{3} < 40$, which are relevant to Refs.~\cite{pellegrino2014,jennewein2016}, only first-order correlations are needed to obtain resonably good agreement with Eq.~(\ref{eq:main_eq_low}). This shows that for these densities, keeping only the first-order correlations is sufficient.

\begin{figure}[!tp]
	\centering
	\includegraphics[width=0.4\textwidth,scale=0.1]{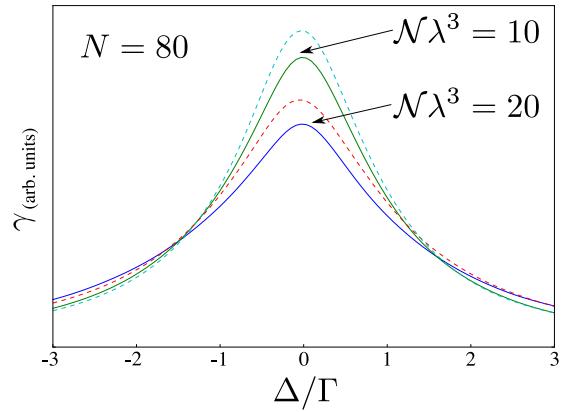}
    \caption{The photon scattering for a spherically symmetric gaussian cloud of $80$ $J_{g} = 2 \rightarrow J_{e} = 3$ atoms. The dotted line corresponds to the results obtained using Eq.~(\ref{eq:ruost_post}), which includes no Zeeman correlations, while the solid line is the calculation that includes the first-order correlations, given by Eq.~(\ref{eq:closed}).}
	\label{fig:gausscompareruo}    
\end{figure}

Understanding the utility of keeping only the first-order correlations, we may apply our calculations to systems of experimental interest, i.e. atoms® with $F_{g} = 2 \rightarrow F_{e} = 3$. In Fig.~\ref{fig:gausscompareruo} we compare the photon scattering lineshape of a spherically symmetric cloud of $80$ atoms using both Eq.~(\ref{eq:ruost_post}), which includes no correlations, and the calculations keeping the first-order correlations, see Eq.~(\ref{eq:closed}). The presence of first-order correlations broadens the lineshape to a non-trivial degree, suppressing the on-resonant scattering in particular. This again indicates the importance of correlations between Zeeman states upon calculating the linehshape. 

\begin{figure}[!tp]
	\centering
	\includegraphics[width=0.4\textwidth,scale=0.1]{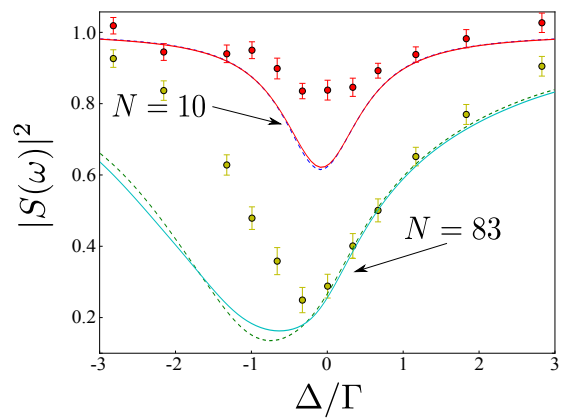}
    \caption{$|S(\omega)|^2$, where $S(\omega)$ is the transfer function defined in the text, versus laser detuning, $\Delta$. The dotted line corresponds to the results obtained using Eq.~(\ref{eq:ruost_post}), which includes no Zeeman correlations, while the solid line is the calculation that includes the first-order correlations, given by Eq.~(\ref{eq:closed}). The solid circles correspond to the experimental data originally reported in Ref. \cite{jennewein2016} and the calculations are conducted for these experimental specifications, as described in the text.}
	\label{fig:transfercompare}    
\end{figure}

Reference \cite{jennewein2016} reports the coherent light transported through a cloud of atoms. They observe the transfer function:
\begin{eqnarray}
S(\omega) &= \nonumber & \frac{(\boldsymbol{E}_{l} + \boldsymbol{E}_{s})\cdot \hat{\boldsymbol{z}}}{\boldsymbol{E}_{l}\cdot \hat{\boldsymbol{z}}} \\
&=& 1 + \frac{\boldsymbol{E}_{s}\cdot \hat{\boldsymbol{z}}}{\boldsymbol{E}_{l}\cdot\hat{\boldsymbol{z}}},
\end{eqnarray}
which is essentially a quantification of the interference between the electric field from the laser, $\boldsymbol{E}_{l}$, and the electric field scattered off of the atoms, $\boldsymbol{E}_{s}$, at the position of the detector. In Fig.~\ref{fig:transfercompare} we compare the results from both Eq.~(\ref{eq:ruost_post}) and the calculations that keep the first-order correlations in Eq.~(\ref{eq:main_eq_low}). This simulation uses the experimental parameters given in Ref.~\cite{jennewein2016}. The cloud now has a fixed size: $\sigma_{x} = 1.5\lambda$, $\sigma_{y} = \sigma_{z} = 0.25\lambda$, which for the clouds of $10$ and $83$ atoms shown, give values of $\lambda^{3}\mathcal{N} \simeq 2.4$ and $\lambda^{3}\mathcal{N} \simeq 19.9$ respectively. Unlike the previous simulations given above, the probe beam is now a linearly polarized Gaussian laser propagating in the $\hat{\boldsymbol{x}}$ direction, with an amplitude:
\begin{equation}
E_{l,z}(\boldsymbol{r}) = \frac{E_{0}}{1 + i\frac{x}{x_{R}}}\exp\Big[ik\frac{y^{2} + z^{2}}{2q(x)}\Big]\exp[ikx],
\end{equation}
where $\frac{1}{q(x)} = \frac{1}{R(x)} + \frac{2i}{k{\rm w}^{2}(x)}$, $x_{R} = k{\rm w}^{2}/2$ is the Rayleigh length, ${\rm w}(x) = {\rm w}\sqrt[•]{1 + x^{2}/x_{R}^{2}}$, $E_{0}$ is the field amplitude, and $R(x) = x + x^{2}_{R}/x$. Where ${\rm w} = 1.54\lambda$ is the waist of the laser. When comparing with the experimental data, one can see that the inclusion of correlations between Zeeman states does not explain the difference between the observations and the calculations. While, similar to experimental observations, the calculations that include Zeeman correlations have a significantly smaller redshift than the results obtained using Eq.~(\ref{eq:ruost_post}), the calculated lineshape is still  non-Lorentzian. This differs from experiment, where the key qualitative experimental observation is that the lineshape is Lorentzian \cite{jennewein2016}. This could indicate that the assumption of an initial, completely uncorrelated, mixture of Zeeman states is incorrect.

\section{Conclusion}

In this work, it is demonstrated that in certain atomic systems, correlations between degenerate Zeeman ground states can have an important effect on the scattered light. This is first shown for 1-D arrays of atoms, where such correlations qualitatively change the photon scattering lineshape of a closely spaced array. It is also shown that for Gaussian clouds of multi-level atoms, the photon scattering rate can be non-trivially changed by these correlations.

In order to simulate these regimes, we developed a set of equations capable of accurately simulating large ensembles of atoms coupled to the electromagnetic field, compared to the full master equation. For $J_{g} = 0$ atoms, this system is identical to the coupled-hamonic oscillator equations commonly used in the literature, allowing Eq.~(\ref{eq:main_eq_low}) to be interpreted as a generalization of this approach. While this set of equations is capable of probing new regimes (see Fig.~\ref{fig:ruolinecompare}), it falls short of the capacity to simulate the large ensembles of atoms present in many experiments. Because of this, an approximation capable of simulating ensembles on the order of hundreds of atoms is developed, see Eq.~(\ref{eq:main_eq_low}). It is then demonstrated that in many systems, this approximation gives nearly identical results to Eq.~(\ref{eq:main_eq_low}), indicating its potential utility in reproducing experimental results in the future. Thus far, the effects of non-degenerate Zeeman structures has remained largely unexplored. In many cases, such atoms are often replaced by two-level atoms with modified dipole-dipole interactions. This work, for the first time, demonstrates that this approach is often inadequate, since more complicated angular momentum structures result in correlations that produce rich physics, and many systems cannot be understood unless this is explicitly considered.

The authors would like to thank Janne Ruostekoski and Antoine Browaeys for informative conversations and data. This material is based upon work supported by the National Science Foundation under Grant No. 1404419-PHY. This research was supported in part through computational
resources provided by Information Technology at Purdue,
West Lafayette, Indiana.

\bibliography{bibtex.bib}

\end{document}